\documentclass[twocolumn,secnumarabic,superscriptaddress,amssymb, nobibnotes, aps, prl]{revtex4-2}

\usepackage{graphicx}  
\graphicspath{{figs/}}
\usepackage{dcolumn}   
\usepackage{bm}        
\usepackage{amssymb}   
\usepackage{amsmath}   
\usepackage{amsthm}    
\usepackage{xcolor}    
\usepackage{tikz}      
\usepackage{ifthen}    
\usepackage{multirow}  
\usepackage{tabu}      
\usepackage{soul}      
\usepackage{subfigure}
\usepackage{xspace}
\usepackage{afterpage}
\usepackage{xr-hyper}
\usepackage[unicode=true,
  linktocpage,
  linkbordercolor={0.5 0.5 1},
  citebordercolor={0.5 1 0.5},
  linkcolor=blue]{hyperref}

\newcommand{\sro}{Sr\ensuremath{_2}RuO\ensuremath{_4}\xspace}
\newcommand{\Tc}{\ensuremath{T_{\mathrm{c}}}\xspace}
\newcommand{\mstar}{\ensuremath{m^{\star}}\xspace}
\newcommand{\lfree}{\ensuremath{l_{\rm{free}}}\xspace}
\newcommand{\wc}{\ensuremath{\omega_{\rm{c}}}\xspace}
\newcommand{\pwave}{\ensuremath{p-}wave\xspace}

\newcommand{\tauq}{\ensuremath{\tau_{\mathrm{q}}}\xspace}

\begin{document}

\title{Quantum Oscillations and the Quasiparticle Properties of Thin Film $\mathrm{Sr}_2\mathrm{RuO}_4$}%

\author{Yawen Fang}
\affiliation{Laboratory of Atomic and Solid State Physics, Cornell University, Ithaca, NY, 14853, USA}
\author{Hari P. Nair}
\affiliation{Department of Materials Science and Engineering, Cornell University, Ithaca, NY, 14853, USA}
\author{Ludi Miao}
\affiliation{Laboratory of Atomic and Solid State Physics, Cornell University, Ithaca, NY, 14853, USA}
\author{Berit Goodge}
\affiliation{School of Applied and Engineering Physics, Cornell University, Ithaca, NY, 14853, USA}
\author{Nathaniel J. Schreiber}
\affiliation{Department of Materials Science and Engineering, Cornell University, Ithaca, NY, 14853, USA}
\author{Jacob P. Ruf}
\affiliation{Laboratory of Atomic and Solid State Physics, Cornell University, Ithaca, NY, 14853, USA}
\author{Lena F. Kourkoutis}
\affiliation{School of Applied and Engineering Physics, Cornell University, Ithaca, NY, 14853, USA}
\affiliation{Kavli Institute at Cornell for Nanoscale Science, Ithaca, New York 14853, USA}
\author{Kyle M. Shen}
\affiliation{Laboratory of Atomic and Solid State Physics, Cornell University, Ithaca, NY, 14853, USA}
\affiliation{Kavli Institute at Cornell for Nanoscale Science, Ithaca, New York 14853, USA}
\author{Darrell G. Schlom}
\affiliation{Department of Materials Science and Engineering, Cornell University, Ithaca, NY, 14853, USA}
\affiliation{Kavli Institute at Cornell for Nanoscale Science, Ithaca, New York 14853, USA}
\affiliation{Leibniz-Institut für Kristallzüchtung, Max-Born-Str. 2, 12489 Berlin, Germany}
\author{B.~J.~Ramshaw}\email[To whom correspondence should be addressed, ]{bradramshaw@cornell.edu}
\affiliation{Laboratory of Atomic and Solid State Physics, Cornell University, Ithaca, NY, 14853, USA}

\begin{abstract}
We measure the Shubnikov-de Haas effect in thin-film \sro grown on an (LaAlO$_3$)$_{0.29}$-(SrAl$_{1/2}$Ta$_{1/2}$O$_3$)$_{0.71}$ substrate. We detect all three known Fermi surfaces and extract the Fermi surface volumes, cyclotron effective masses, and quantum lifetimes. We show that the electronic structure is nearly identical to that of single-crystal \sro, and that the quasiparticle lifetime is consistent with the \Tc of comparably clean, single-crystal \sro. Unlike single-crystal \sro, where the quantum and transport lifetimes are roughly equal, we find that the transport lifetime is $1.3\pm0.1$ times longer than the quantum lifetime. This may suggest that extended (rather than point) defects could be the dominant source of quasiparticle scattering in these films. To test this hypothesis, we perform cross-sectional scanning transmission electron microscopy and find that out-of-phase boundaries extending the entire thickness of the film occur with a density that is consistent with the quantum mean free path. The long quasiparticle lifetimes make these films ideal for studying the unconventional superconducting state in \sro through the fabrication of devices---such as planar tunnel junctions and  superconducting quantum interference devices.
\end{abstract}

\maketitle

\section{Introduction}

 \sro was long thought to be a \pwave superconductor, but recently revised nuclear magnetic resonance (NMR) measurements find a substantial decrease in the Knight shift across the superconducting transition temperature \Tc \cite{PustogowNat2019}, essentially ruling out all spin-triplet pairing states. While the details of the superconducting state are far from settled \cite{kivelson2020proposal,ghosh2020thermodynamic}, it appears that single-crystal \sro is not a $p_x + i p_y$ superconductor. It may be possible, however, that multiple superconducting order parameters lie nearby in energy \cite{raghu2010hidden,zhang2018superconducting}, suggesting that the application of the right tuning parameter could push \sro into a \pwave state. Uniaxial strain along the $[100]$ direction has been shown to strongly enhance \Tc \cite{hicks2014strong}, and while there is no signature of \pwave superconductivity under strain in single-crystals, these measurements suggest that strain is a good parameter for manipulating the superconducting state of \sro.

Unlike single-crystal strain experiments, which are necessarily uniaxial or hydrostatic, thin films can be strained biaxially through substrate engineering. For example, using five different substrates, \citet{burganov2016strain} showed that the $\gamma$ Fermi surface sheet of \sro (and the closely related compound Ba$_2$RuO$_4$) can be driven through the Brillouin zone boundary. These films, however, were not superconducting, and for decades the growth of superconducting thin-film \sro has been a major challenge in the oxide thin film community \cite{madhavan1996a}. The difficulty stems from the extreme sensitivity of \sro to even minute levels of disorder---single crystals with greater than 1 $\mu$Ohm-cm residual resistivity do not superconduct \cite{MackenziePRL1998}. With the advent of \sro films that are clean enough to show superconductivity on many different substrates \cite{nair2018demystifying,uchida2017molecular,garcia2020pair}, it is worth investigating whether the superconductivity is a product of film quality, substrate strain, or both, how the quasiparticle properties are modified by the substrate, and what types of defects might be limiting the quasiparticle mean free path.

\section{Experiment}
\begin{figure*}[t!]
\includegraphics[width=1.5\columnwidth]{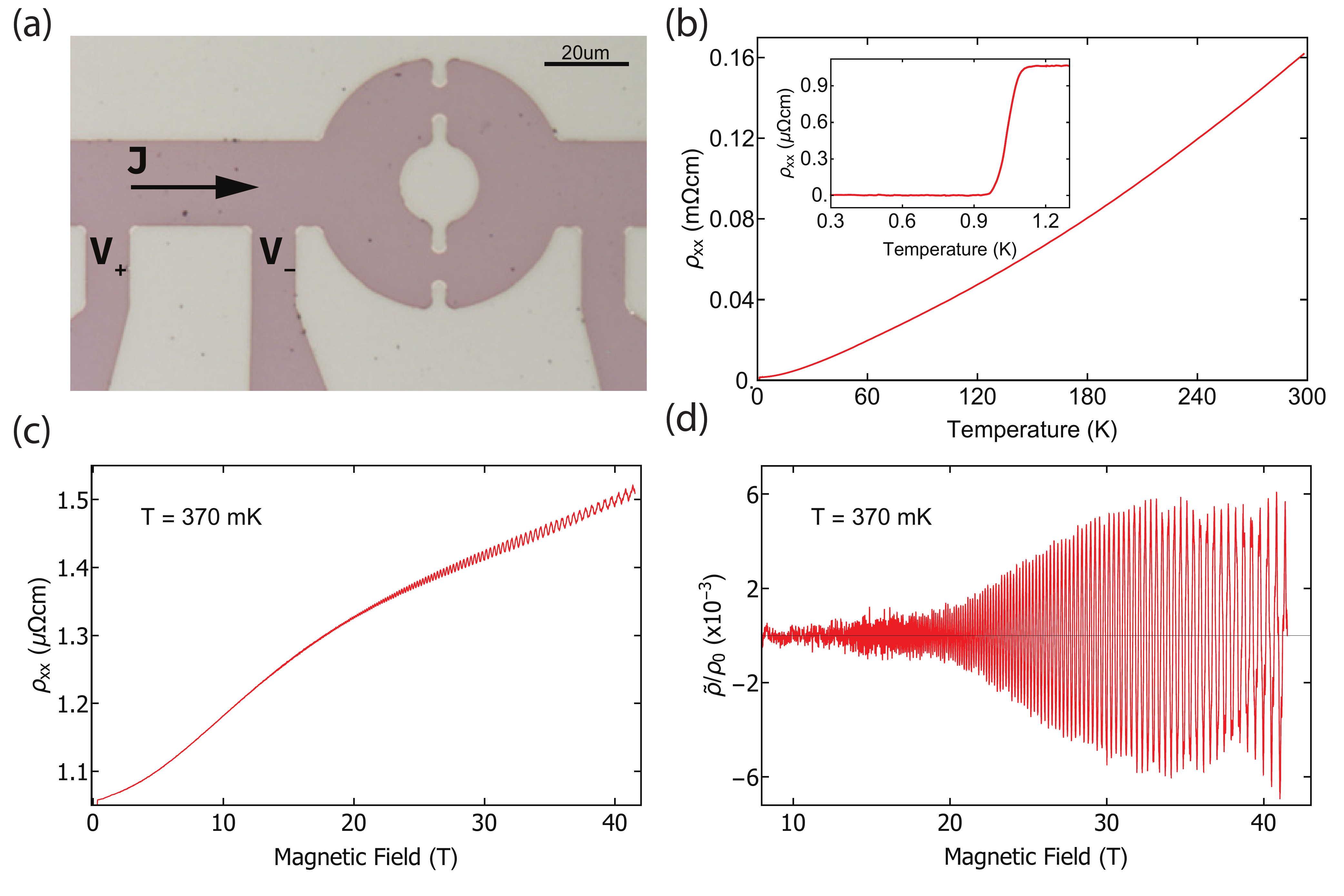}%
\caption{\textbf{Shubnikov-de Haas oscillations in $\mathrm{Sr}_2\mathrm{RuO}_4$ on LSAT}. (a) The \sro transport device patterned from a thin film of \sro grown on an LSAT substrate. The structure to the right of the voltage contacts was not part of this experiment. (b) Resistivity as a function temperature: the inset shows the superconducting transition with a midpoint of $T = 1.05$~K. (c) Resistivity as a function of magnetic field. (d) The normalized oscillatory component of the resistivity.}%
\label{fig:device}
\end{figure*}

A 100 nm thick film of \sro was grown via molecular-beam epitaxy on an (LaAlO$_3$)$_{0.29}$-(SrAl$_{1/2}$Ta$_{1/2}$O$_3$)$_{0.71}$ (LSAT) substrate with the tetragonal $c$-axis perpendicular to the substrate surface. This substrate imposes a 0.045\% tensile strain (a dilation of the tetragonal unit cell) at low temperature. The growth procedure is described in Refs. \cite{nair2018demystifying,mark2001suppression,mark2003defect}. Devices for measuring electrical resistivity in the $ab$-plane were fabricated using standard photolithography techniques and ion milling (see \autoref{fig:device}a). Devices were contacted by sputtering a 5 nm titanium adhesion layer followed by 25 nm of platinum. The temperature dependence of the in-plane resistivity, $\rho_{\rm xx}$, reveals a high-quality device, with a RRR of 106 and a superconducting \Tc of 1.05 K (RRR is defined as $R(298 K)/R(T_c)$, with \Tc measured at the midpoint of the transition---see \autoref{fig:device}b.) 

Quantum oscillation measurements were performed in the 41 T resistive magnet at the National High Magnetic Field Lab in Tallahassee. The magnetic field was oriented parallel to the crystallographic c-axis—perpendicular to the plane of the device. The temperature was controlled at fixed intervals between 370 mK and 2.2 K in a helium-3 cryostat. The sample resistance was measured in a standard 4-point contact geometry using a Stanford Research 860 lock-in amplifier, with a Stanford Research CS 580 Voltage Controlled Current Source and a Stanford Research SR560 low-noise preamplifier. The current through the sample was $I_{\rm{pp}}= 150 ~\mu$A. The magnetic field was swept at a rate of 0.2 T/min, and the time constant of the lock-in amplifier was set to 1 s. The slow sweep rate ensured that the high-frequency oscillations were not washed out by the time constant of the lock-in amplifier.

Figure 1(c) presents the magnetic field dependence of the resistivity at 370 mK. The total resistivity $\rho(B)$ is composed of two parts: a non-oscillatory background $\rho_0(B)$, which we obtain by fitting a smooth polynomial to the data, and an oscillatory component $\widetilde{\rho}(B)$. The oscillatory fraction of the resistivity, $\widetilde{\rho}/\rho_0$ \cite{shoenberg1984magnetic}, is given by 
\begin{equation}
\frac{\widetilde{\rho}}{\rho_0}=(\frac{{\rho}}{\rho_0}-1)
\end{equation}
and is plotted in \autoref{fig:device}d. Shubnikov–de Haas oscillations are clearly visible above 15 T, with multiple frequencies visible above 35 T (see \autoref{fig:oscillations}a and the supplementary information
for details about the data analysis).

\section{Analysis}

\begin{figure*}[t!]
\includegraphics[width=1.5\columnwidth]{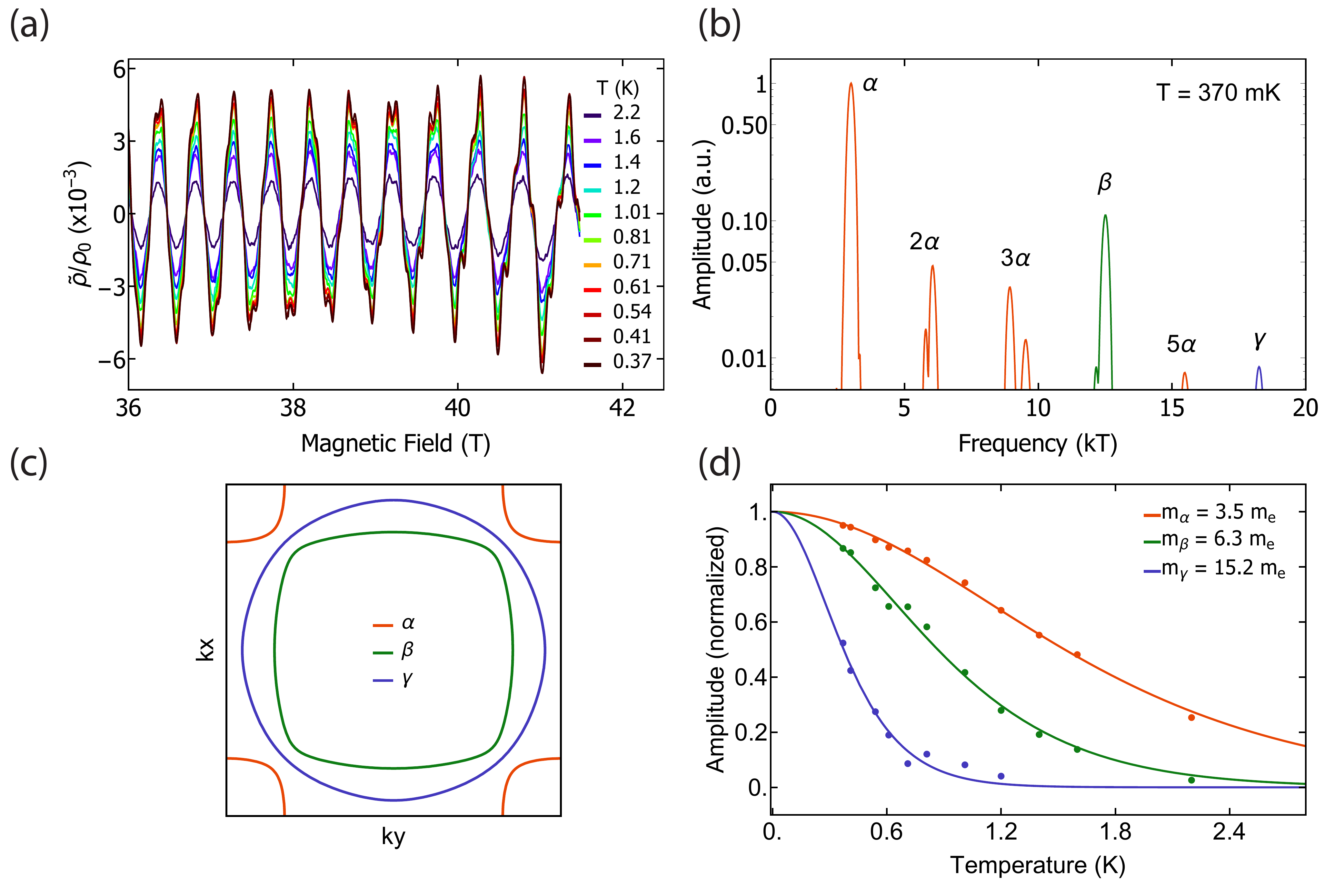}%
\caption{\textbf{Analysis of the Shubnikov-de Haas oscillations.} (a) Temperature dependence of the Shubnikov-de Hass oscillations in the field range from 36 to 41.5 T. Additional frequency components---beyond the 3 kT frequency of the $\alpha$ pocket---become clearly visible at higher field. (b) Fast Fourier transform (FFT) of the 370 mK data. The primary frequencies corresponding to the $\alpha$, $\beta$, and $\gamma$ Fermi surfaces (shown in (c)) are indicated, as well as the higher harmonics of the $\alpha$ surface. (d) Temperature-dependent oscillation amplitude, with a fit to \autoref{eq:lk} for the $\alpha$, $\beta$, and $\gamma$ pockets. The analysis is done with data between 32 T and 41.5 T and from 370 mK to 2.2 K for $\alpha$ and $\beta$ pockets, and between 35 T and 41.5 T and from 370 mK to 1.2 K for the $\gamma$ pocket.}
\label{fig:oscillations}
\end{figure*}

We analyze the temperature and field dependence of the Shubnikov-de Haas oscillations to determine the Fermi surface area, the quasiparticle effective mass, and the quasiparticle mean free path, for all three sheets of the Fermi surface. The fast Fourier transform (FFT) of $\widetilde{\rho}/\rho_0$, shown in \autoref{fig:oscillations}b, has clear contributions from all three known pieces of Fermi surface, labeled $\alpha$, $\beta$, and $\gamma$, in accordance with previous studies \cite{bergemann2003quasi}. Harmonics from the $\alpha$ pocket are visible up to the fifth order---another indication of high sample quality, as harmonic amplitude is dampened out exponentially with increasing harmonic number. 

The Fermi surface area $A_k$ is obtained from the quantum oscillation frequency $F$ through the Onsager relation $A_k = (2\pi e/\hbar)F$. LSAT substrates apply a relatively small amount of tensile strain on the \sro films, and thus we expect the Fermi surface area of \sro films to be close to what is measured in single-crystal \sro. This is what we observe: within our experimental resolution, the three Fermi surfaces of \sro grown on LSAT have the same area as those reported for single crystals (a comparison is shown in \autoref{tab:comp}). This is consistent with the relatively small, 0.045\% $A_{1g}$ strain imposed by LSAT to the film at low temperature. 

The quasiparticle effective mass \mstar is obtained from the temperature dependence of the quantum oscillation amplitude. As the thermal energy ($k_B T$) becomes comparable to the cyclotron energy ($\hbar \omega_{\mathrm{c}}$, where $\omega_{\mathrm{c}} = \frac{eB}{\mstar}$), the oscillations are damped with the form 
\begin{equation}
R_{\rm{T}} = \left(2 \pi^2 \frac{k_{\mathrm{B}} T}{\hbar \omega_{\mathrm{c}} }\right)/\sinh\left(2 \pi^2 \frac{k_{\mathrm{B}} T}{\hbar \omega_{\mathrm{c}} }\right).
\label{eq:lk}
\end{equation}
Figure 2(d) shows fits of the FFT amplitude to \autoref{eq:lk}. The effective mass for the $\alpha$, $\beta$, and $\gamma$ sheets are found to be $m^{\star}_\alpha = $ 3.5$\pm$0.1 $m_e$, $m^{\star}_\beta = $ 6.3$\pm$0.2 $m_e$, and $m^{\star}_\gamma = $ 15.2$\pm$1.3 $m_e$, where $m_e$ is the bare electron mass. These masses are consistent with those found in single-crystal \sro (see \autoref{tab:comp} for a comparison), which is also consistent with the similarity in measured Fermi surface areas. The consistency between these Fermi surface parameters suggests that films of \sro grown on LSAT are close electronic analogs of the bulk material.

The remaining quantity to be determined is the quasiparticle quantum lifetime $\tau_q$. As the inverse of the time between scattering events becomes comparable to the cyclotron frequency \wc (or the Landau level widths becomes comparable to their separation), the oscillations are damped as
\begin{equation}
 R_{\rm{D}}=e^{-\frac{\pi }{\wc\tau_q}}.
\label{eq:ding}
\end{equation}
The lifetime can be converted to a mean free path via $\tau_q = \lfree/v_{\rm{F}}$, where the Fermi velocity $v_{\rm{F}}$ is determined by the measured Fermi surface area and effective mass. 

The quantum lifetime is more challenging to extract than the cyclotron mass and Fermi surface area for two reasons. First, the $\alpha$ pocket dominates the raw oscillatory signal, making it impossible to fit \autoref{eq:ding} directly to the data for the $\beta$ and $\gamma$ bands. Second, the presence of interlayer coupling leads to a beat-like structure in the data rather than a pure exponential envelope \cite{bergemann2003quasi}. We solve the first problem by Fourier-filtering the data over frequency ranges that only include one FS component at a time. We solve the second by fitting the data over as broad a field-range as possible and by comparing the results of two different analyses to check for consistency. 

We first extract the quantum lifetime with a Dingle plot. The Shubnikov-de Haas oscillation amplitude for a quasi-2D Fermi surface is given by 
\begin{equation}
\frac{\widetilde{\rho}}{\rho_0} \propto	 R_T R_D \cos\left(\frac{2 \pi F}{B}\right),
\label{eq:oscfull}
\end{equation}
where $R_T$ is given by \autoref{eq:lk} and $R_D$ is given by \autoref{eq:ding}. We first find the peaks of the oscillations in $\left|\widetilde{\rho}/\rho_0\right|$, divide the peak amplitude by $R_T$, and then plot the absolute value of this quantity on a log scale as a function of 1/$B$; these plots are shown in \autoref{fig:dingle1}. With \mstar determined from the temperature dependence, the quantum lifetimes can be immediately obtained from the slope of this plot. 

The second method we use to determine the quantum lifetime is to directly fit the oscillations with \autoref{eq:oscfull}---these plots are shown in \autoref{fig:dingle2}. This has the advantage over the first technique of making use of the full data set, but as \autoref{eq:oscfull} ignores the interlayer dispersion (which cannot be clearly resolved over this field range), it can become contaminated by the beat structure.

\begin{table*}[t!]
\caption{A comparison of the Fermi surface parameters extracted for a \sro film grown on LSAT and those obtained from de Haas–van Alphen measurements on single-crystal \sro \cite{bergemann1999quantum,bergemann2001normal} (\tauq for single-crystals comes from \citet{mackenzie1996quantum}.) The transport lifetime $\tau_t$ is calculated from the value of $\rho_{\rm xx}(\Tc)$ assuming that the ratio $\tau_t/\tau_q$ is the same for all sheets of Fermi surface. The details of the estimation of uncertainties are described in the supplementary information.}
  \begin{ruledtabular}

    \begin{tabular}{l l l l l} 
        Fermi surface & Frequency ( T ) &  Effective mass ( $m_e$ ) &\vtop{\hbox{\strut $\tau_{q}$ ( ps )}\hbox{\strut (quantum lifetime)}} &\vtop{\hbox{\strut $\tau_{t}$ ( ps )}\hbox{\strut (transport lifetime)}} \\
        \hline
       $\alpha$ ( film )  & 3079 $\pm$ 99 & 3.5 $\pm$ 0.1  & 1.07 $\pm$ 0.08 & 1.43 $\pm$ 0.10 \\
	   $\alpha$ ( single crystal )  &3010 $\pm$ 80 & 3.4   $\pm$ 0.1  & 2.0 & \\ \hline
	  $\beta$ ( film ) & 12510 $\pm$ 108 & 6.3 $\pm$ 0.2  & 0.66 $\pm$ 0.03 & 0.88 $\pm$0.04 \\
	  $\beta$ ( single crystal )  &12730 $\pm$ 150 & 6.8 $\pm$ 0.2 & 1.7 & \\ \hline
	   $\gamma$ ( film ) & $18259 \pm 195$ & 15.2 $\pm$ 1.3  & 0.71 $\pm$ 0.13 & 0.95 $\pm$ 0.17\\
	  $\gamma$ ( single crystal )  &18570 $\pm$ 70 & 14.0 $\pm$ 2.0 & 2.4 & \\
    \end{tabular}
  \end{ruledtabular}

    \label{tab:comp} 
\end{table*}

The two methods give similar estimates for $\tau_q$. We take the average of the two results and estimate the uncertainty as the difference between them. The quantum lifetimes of the $\alpha$, $\beta$, and $\gamma$ sheets are 1.07$\pm 0.08$ ps, 0.66$\pm 0.03$ ps, and 0.71$\pm 0.13$ ps, respectively. As the $\gamma$ pocket was only observed above 35 T, there is certainly a larger systematic uncertainty associated with this lifetime than we are able to account for with this method. Converting the three lifetimes to mean free paths yields 108 nm, 75 nm, and 40 nm, for the $\alpha$, $\beta$, and $\gamma$ Fermi surfaces, respectively. These values can be compared with those obtained from single crystals: \citet{mackenzie1996quantum} report Dingle temperatures that convert to mean free paths of 210 nm, 176 nm, and 130 nm, for the $\alpha$, $\beta$, and $\gamma$ Fermi surfaces, respectively. While the single-crystal values are somewhat longer than those from our film, the progression of the longest mean free path on the $\alpha$ pocket to the shortest on the $\gamma$ pocket is consistent (note that subsequent generations of single-crystal \sro have even longer mean free paths \cite{bergemann2003quasi}.)

The transport lifetime $\tau_t$---a quantity related to, but distinct from, the quantum lifetime---can be extracted from the absolute value of the resistivity $\rho_{\rm xx}$ now that the Fermi surface areas and effective masses are known. We start with a tight-binding model of \sro's band structure, adjust the tight-binding parameters so that the Fermi surface areas and effective masses match the values measured for our sample, and then solve the Boltzmann transport equation using Chambers' solution \cite{chambers1952kinetic} (details are given in the supplementary information.) As there are three transport lifetimes---one for each Fermi surface---but only one value of $\rho_{\rm xx}$ to fit, we make the simplifying assumption that the ratio of $\tau_t$ to $\tau_q$ is the same for all sheets of Fermi surface. We adjust this ratio until the calculated resistivity matches the measured residual resistivity, $\rho_{\rm xx}(\Tc)$=1 $\mu\Omega\cdot$cm. We find that $\tau_t/\tau_q = 1.3\pm0.1$. This translates to transport mean free paths of 140 nm, 97 nm, and 52 nm, for the $\alpha$, $\beta$, and $\gamma$ Fermi surfaces, respectively. Because $\tau_q$ is a lower bound on $\tau_t$, the ratio $\tau_t/\tau_q$ must be at least 1. If we relax the constraint that $\tau_t/\tau_q$ is the same for all Fermi surfaces, then at most two Fermi surfaces could have $\tau_t/\tau_q = 1$ and the third would have $\tau_t/\tau_q$ greater than 1.3 (the exact value depends on which Fermi surface is chosen). Without further microscopic justification for why $\tau_t/\tau_q$ might be different on different Fermi surfaces, the assumption that this ratio is the same for all Fermi surfaces is the simplest one that we can make.
\section{Discussion}

The transport lifetimes we measure here approach those of clean-limit, single-crystal \sro \cite{MackenziePRL1998}. They are also comparable to what was reported in some of the earliest quantum oscillation measurements of single-crystal \sro \cite{mackenzie_calculation_1996}. To put our measured mean free path of over 100 nm in context with other oxide thin film superconductors, a useful comparison can be made with Pr$_{2-x}$Ce$_x$CuO$_{4\pm\delta}$, whose crystal structure is very simliar to that of \sro. High-field quantum oscillation studies on Pr$_{2-x}$Ce$_x$CuO$_{4\pm\delta}$ have measured the mean free path to be only 6 nm \cite{breznay2016shubnikov}, highlighting the extremely high quality of our \sro films. 

Long mean free paths are crucial for observing the intrinsic \Tc of \sro: \citet{MackenziePRL1998} found that 90 nm is the critical transport mean free path for superconductivity in \sro---any shorter and the material does not superconduct; any longer and the \Tc rises rapidly to saturate at $\approx 1.5$~K. We find mean free paths more than twice this length on the $\alpha$ and $\beta$ bands, which are thought to dominate the superconductivity in \sro \cite{sharma2019momentum}, and which is consistent with a \Tc of 1.05 K for single-crystal \sro \cite{MackenziePRL1998}.

The difference between the measured quantum and transport lifetimes may offer a clue as to what is the dominate scattering mechanism in these films. The quantum lifetime is the average time a quasiparticle spends in a momentum eigenstate before scattering. The transport lifetime is the average time between scattering events that relax the quasiparticle momentum distribution function. When scattering is isotropic, as it is for point-scatterers, $\tau_q = \tau_t$. For extended defects, the transport lifetime is generally longer than its quantum counterpart: extended defects contribute more forward-scattering events that do not alter the momentum distribution function of the quasiparticles but \textit{do} decohere the quasiparticle wavefunctions. This was studied systematically in AlGaN/GaN heterostructures, where small-angle scattering from dislocations reduces the quantum lifetime by up to a factor of 20 below the transport lifetime \cite{manfra_quantum_2004}.  While a ratio of $1.3\pm0.1$ is not nearly as compelling as a ratio of 20, we were nevertheless motivated to study the microscopic nature of the defects in this film.

\begin{figure*}[t!]
\includegraphics[width=1.5\columnwidth]{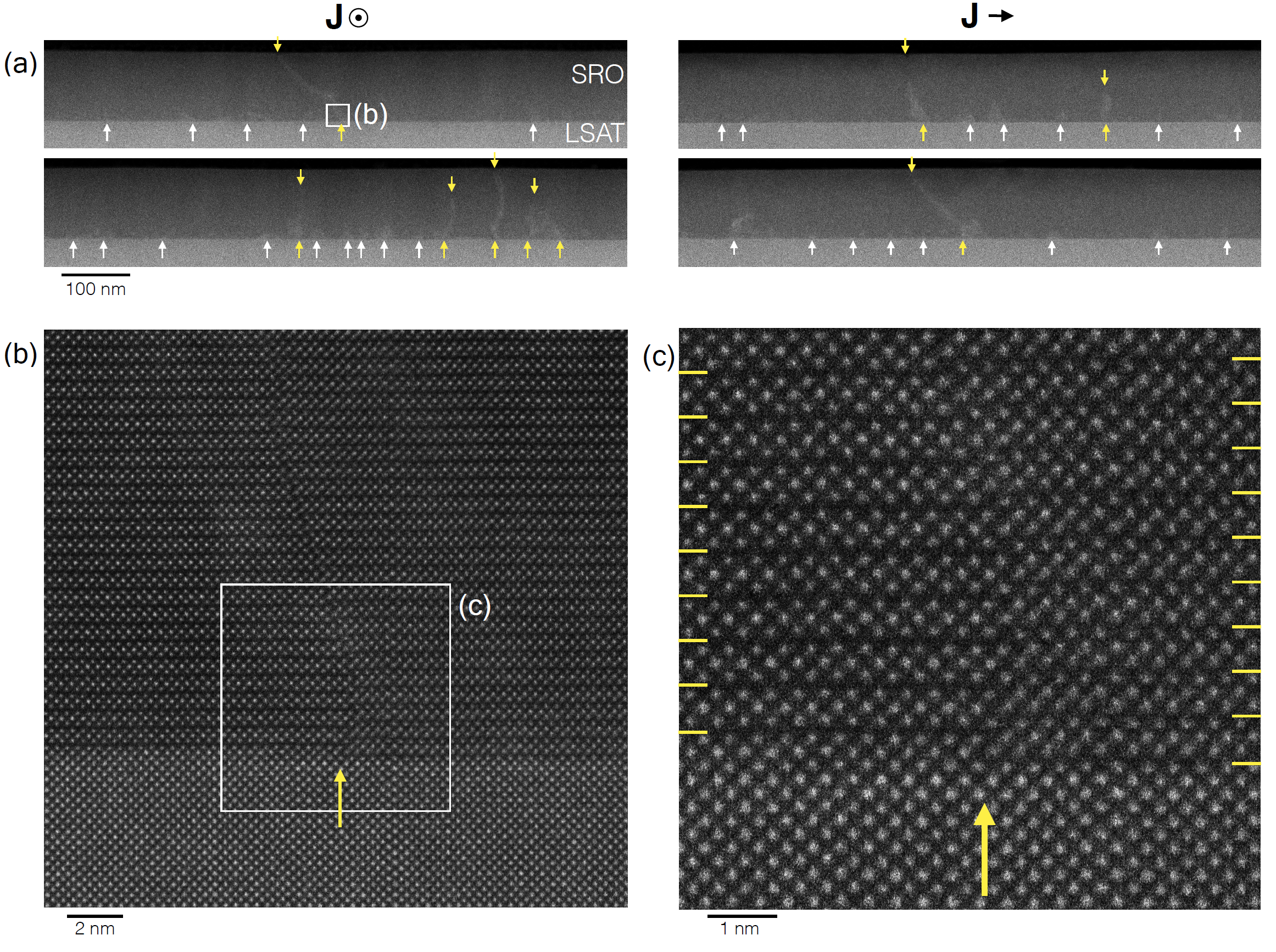}%
\caption{\textbf{Defect characterization by cross-sectional STEM.} Atomic-resolution medium angle annular dark field scanning transmission electron microscopy (MAADF-STEM) images for cross sections perpendicular (a) and parallel (b) to the current direction in \autoref{fig:device}a. Extended lattice defects, such as out-of-phase boundaries, are indicated with arrows. Many defects terminate near the interface of the film (white arrows), while some are observed to extend through more than half the film thickness (yellow arrows). (c) Nearly all out-of-phase boundaries can be traced to nucleate at step edges (yellow arrow) in the LSAT substrate surface, as seen in this atomic-resolution image of a defect nucleation from the region marked by the white box in (a). (d) High-magnification inset of the area marked by the white box in (c) shows how \sro layers growing near a single unit cell LSAT step edge meet at a vertical defect due to the vertical offset of SrO planes on either side of the step edge. }
\label{fig:stem}
\end{figure*}

To investigate the character and density of extended defects in our samples, we performed cross-sectional STEM on lamellas cut from the contact region of the device shown in \autoref{fig:device}a. \autoref{fig:stem}a and b show representative cross sections cut parallel and perpendicular to direction of the applied current, respectively (more images are shown in the SI.) Extended defects---predominantly out-of-phase boundaries---are visible as regions of lighter contrast in these images and are present with a density of approximately 1 every 100 to 200 nm. This density is consistent with the quantum mean free path we extract from the quantum oscillations. The longer transport mean free path, as compared to the quantum mean free path, is consistent with the predominantly small-angle scattering that results from the large spatial extent of these defects \cite{manfra_quantum_2004}. \autoref{fig:stem}c and d show that step-edges in the LSAT substrate are predominant nucleation sites for these defects. This suggests that future improvements in \sro film quality should focus on reducing the density of step edges through substrate surface preparation. 

The extreme sensitivity of \Tc to disorder in \sro raises the question of whether the relatively high \Tc observed in thin films can be attributed to film cleanliness or whether the presence of the substrate significantly modifies the electronic structure and thus \Tc. The Fermi surface areas and quasiparticle effective masses we measure for \sro grown on LSAT are the same as those found in single-crystal \sro to within our measurement uncertainty. This suggests that modifications to the electronic structure, such as an enhanced density of states (proportional to \mstar in two-dimensional materials) due to substrate strain pushing the Fermi surface toward the van Hove point, are not responsible for the relatively high \Tc (1.05 K) observed in these films grown on the commercial perovskite substrate that is best lattice matched to \sro \cite{schlom2014elastic}. It will be interesting to see how the electronic structure, mean free path, and perhaps even superconducting order parameter symmetry are modified when commensurately strained \sro films are grown on other substrates, where the \Tc can be as high as 1.8 K and strain is undoubtedly playing a larger role \cite{nair2018demystifying}.

\section{Acknowledgments} 

B.J.R and Y.F. acknowledge support from the National Science Foundation under grant no. DMR-1752784. A portion of this work was performed at the National High Magnetic Field Laboratory, which is supported by the National Science Foundation Cooperative Agreement No. DMR-1644779 and the State of Florida. H.P.N., L.M., N.J.S., B.H.G, L.F.K, K.M.S, and D.G.S. acknowledge support from the National Science Foundation (Platform for the Accelerated Realization, Analysis and Discovery of Interface Materials (PARADIM)) under Cooperative Agreement No. DMR-1539918. N.J.S. acknowledges support from the National Science Foundation Graduate Research Fellowship Program under Grant No. DGE-1650441. This research is funded in part by the Gordon and Betty Moore Foundation's EPiQS Initiative through Grant Nos. GBMF3850 and and GBMF9073 to Cornell University. This work made use of the Cornell Center for Materials Research (CCMR) Shared Facilities, which are supported through the NSF MRSEC Program (No. DMR-1719875). Substrate preparation was performed in part at the Cornell NanoScale Facility, a member of the National Nanotechnology Coordinated Infrastructure (NNCI), which is supported by the NSF (Grant No. NNCI-2025233). The FEI Titan Themis 300 was acquired through No. NSF-MRI-1429155, with additional support from Cornell University, the Weill Institute, and the Kavli Institute at Cornell. The Thermo Fisher Helios G4 UX FIB was acquired with support by NSF No. DMR-1539918. K.M.S acknowledges support from NSF DMR-1709255 and Air Force Office of Scientific Research Grant No. FA9550-15-1-0474.

\newpage
\onecolumngrid

\renewcommand{\thefigure}{S\arabic{figure}}

\section{Materials and Methods}
\subsubsection{A. Film synthesis by MBE}
The \sro thin film was grown in a Veeco Gen10 molecular-beam epitaxy (MBE) system on a (LaAlO$_3$)$_{0.29}$-(SrAl$_{1/2}$Ta$_{1/2}$O$_3$)$_{0.71}$ (LSAT) substrate from CrysTec GmbH. The substrate used for the growth was screened to have a miscut of less than $0.05^{\circ}$, which is important to reduce the formation of out-of-phase boundaries \cite{mark2001suppression,mark2003defect,nair2018demystifying}.The films were grown at a substrate temperate of 810 $^{\circ}$C as measured using an optical pyrometer operating at 1550 nm. Elemental strontium (99.99\% purity) and elemental ruthenium (99.99\% purity) evaporated from a low-temperature effusion cell and a Telemark electron beam evaporator, respectively, were used for growing the \sro film. The films were grown with a strontium flux of $2.6 \times 10^{13}$ atoms$\cdot$cm$^{-2}$s$^{-1}$ and a ruthenium flux of $1.8 \times 10^{13}$ atoms$\cdot$cm$^{-2}$s$^{-1}$ in a background of distilled ozone ($\sim$80\% O$_3$ + 20\% O$_2$ made from oxygen gas with 99.994\% purity.) The background oxidant pressure during growth was $3\times10^{-6}$ Torr. At the end of the growth the strontium and ruthenium shutters were closed simultaneously, and the sample was cooled down to below 250 $^{\circ}$C in a background pressure of distilled ozone of $1\times10^{-6}$ Torr. Further details of the adsorption-controlled growth conditions for the growth of \sro thin films by MBE can be found elsewhere \cite{nair2018demystifying}.

\subsubsection{B. Device fabrication}
The \sro films grown by MBE were subsequently patterned into devices for transport measurements with platinum contacts using standard photolithography, sputter deposition, and ion milling techniques. First, the Pt contact geometry was defined using photolithography. Next, 25 nm of platinum, with 5 nm titanium adhesion layer, was sputtered onto the \sro film with an AJA sputtering tool, followed by a standard lift-off processes. A second photolithography step was used to define the transport device geometry, followed by ion milling with an AJA ion mill to remove the excess \sro film. 

\subsubsection{C. Cross sectional STEM imaging}
Cross-sectional STEM specimens were prepared using the standard focused ion beam (FIB) lift-out process on Thermo Scientific Helios G4 UX FIB. Medium-angle annular dark-field scanning transmission electron microscopy (MAADF-STEM) images were acquired on an aberration-corrected FEI Titan Themis operating at 300 keV with a probe convergence semi-angle of 30 mrad and inner and outer collection angles of 46 and 230 mrad, respectively.

\section{Background Remove and Data Processing}
\begin{figure}[h!]
\includegraphics[width=0.8\columnwidth]{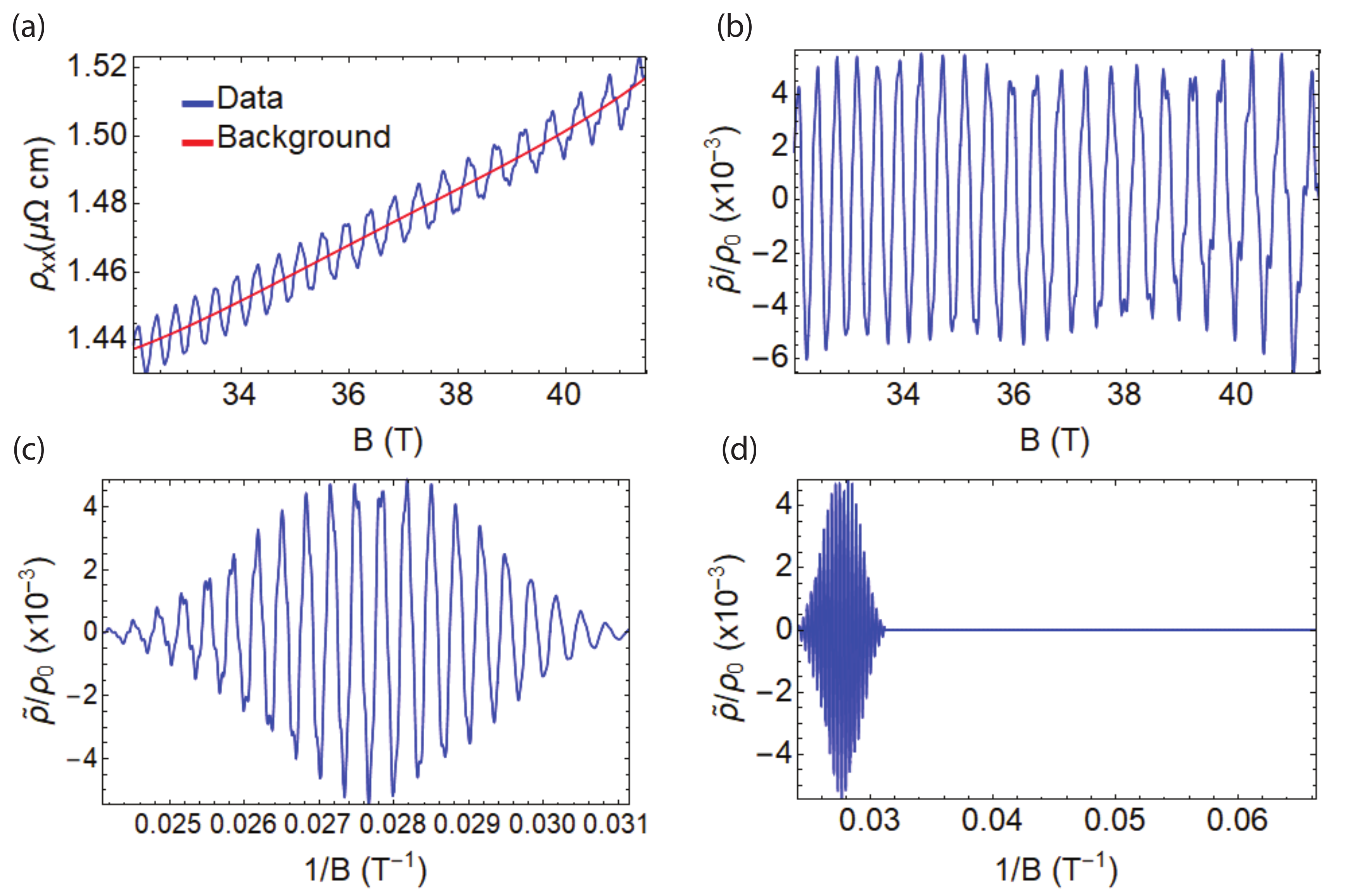}%
\caption{\textbf{Data processing prior to Fourier transforming.}
(a) Resistivity as a function of magnetic field at 370 mK, from 32 T to 41.5 T. (b) The normalized oscillatory component of the resistivity. (c) The normalized resistivity plotted versus inverse field and multiplied by Kaiser-Bessel window with $\alpha = 1.8$. (d) The data from (c) is finally padded with zeroes before Fourier transforming, the result of which is shown in Figure 2b of the main text.
 }
\label{fig:fourier}
\end{figure}

Additional information about the data processing prior to Fourier transforming is given in the caption to \autoref{fig:fourier}. Because all components of the oscillatory signal vary much faster than the background resistivity, the amplitudes extracted from the Fourier transforms are independent of the order of polynomial used for the background. A Kaiser-Bessel window is used to prevent high-frequency artifacts in the FFT, and zero-padding is used for asthetic purposes to smooth the peaks. 

\section{Extraction of the Quantum Lifetime}
Shubnikov-de Haas oscillations in a quasi-2D material are described by 
\begin{equation}
\frac{\widetilde{\rho}}{\rho_0} \propto	 R_T R_D \cos\left(\frac{2 \pi F}{B}\right),
\label{eq:oscfulls}
\end{equation}
where $F$ is the quantum oscillation frequency, the thermal damping factor $R_T$ is given by
\begin{equation}
R_T = \left(2 \pi^2 \frac{k_{\mathrm{B}} T}{\hbar \omega_{\mathrm{c}} }\right)/\sinh\left(2 \pi^2 \frac{k_{\mathrm{B}} T}{\hbar \omega_{\mathrm{c}} }\right),
\label{eq:rt}
\end{equation} 
where $\omega_{\mathrm{c}} = \frac{eB}{\mstar}$, and the damping due to quasiparticle scattering, $R_D$, is given by 
\begin{equation}
R_D = e^{-\frac{\pi }{\wc\tau_{\mathrm{q}}}}.
\label{eq:rd}
\end{equation}
With \mstar entirely determined by fitting the temperature-dependent oscillation amplitude to \autoref{eq:rt}, the quantum lifetime \tauq can be determined by fitting the oscillation envelope to \autoref{eq:rd}.

To fit the $\alpha$, $\beta$, and $\gamma$ sheets separately, we Fourier transform $\frac{\widetilde{\rho}}{\rho_0}$, as shown in Figure 2b of the main text, and then band-pass filter the spectrum to include only one frequency component at a time. We then inverse-Fourier transform the spectrum to recover $\left(\frac{\widetilde{\rho}}{\rho_0}\right)_i$, where $i = \alpha, \beta,$ and $\gamma$. The results are shown in \autoref{fig:dingle2}. Strictly speaking, one should convert from resistivity to conductivity before performing this procedure as conductivities, not resistivity, are additive. However, as $\widetilde{\rho}$ is only a tiny fraction of $\rho_0$, the end result is equivalent (up to an overall minus sign.)

%

\subsection{Method 1: Dingle Plots}
\begin{figure}[h!]
\includegraphics[width=1\columnwidth]{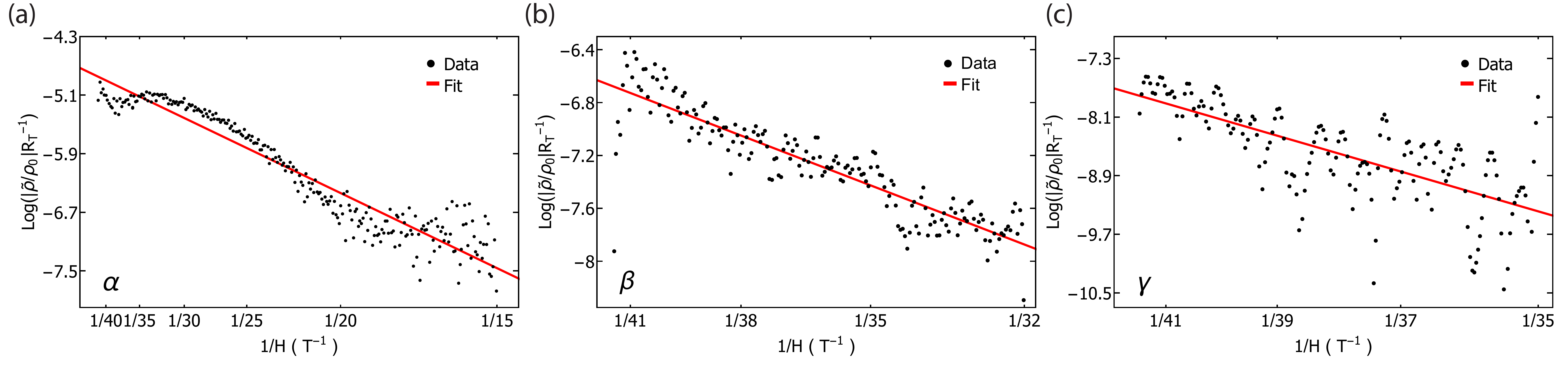}%
\caption{\textbf{Quantum lifetime extraction method 2.}
The black points are the peaks in the absolute value of the oscillatory component for the $\alpha$ (a), $\beta$ (b), and $\gamma$ (c) sheets of Fermi surface. The reds lines are fits to a straight line, yielding quantum scattering times of 1.03 ps, 0.67 ps, and 0.77 ps, for the $\alpha$, $\beta$, and $\gamma$ sheets, respectively.
 }
\label{fig:dingle1}
\end{figure}

\begin{figure}[h!]
\includegraphics[width=1\columnwidth]{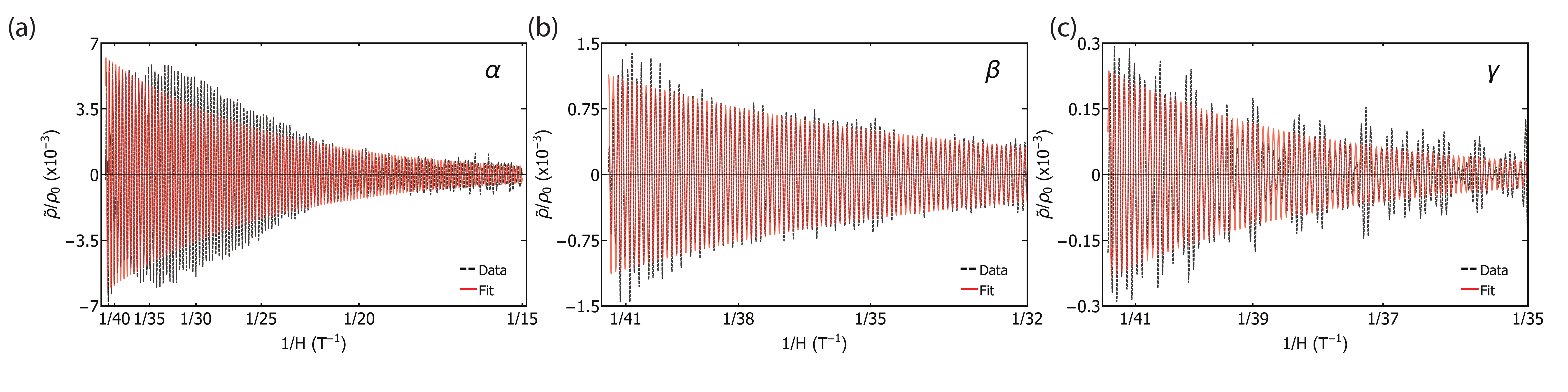}%
\caption{\textbf{Quantum lifetime extraction method 1.} The dashed black lines are the Fourier filtered data for the $\alpha$ (a), $\beta$ (b), and $\gamma$ (c) sheets of Fermi surface. The reds lines are fits to \autoref{eq:oscfulls}, which yield quantum scattering times $\tau_q$ of 1.11 ps, 0.65 ps, and 0.64 ps, for the $\alpha$, $\beta$, and $\gamma$ sheets, respectively.}
\label{fig:dingle2}
\end{figure}

First we extract the quantum lifetimes through what are known as ``Dingle plots.'' For each Fermi surface, we take the absolute value of $\left(\frac{\widetilde{\rho}}{\rho_0}\right)_i$, find all of the peaks, and divide the peak amplitude by the $R_T$ factor for that Fermi surface. This quantity is shown in \autoref{fig:dingle1}. The quantum lifetime 1/$\tau_q$ can be obtained from the slope of the plot, and we find lifetimes of 1.03 ps, 0.67 ps and 0.77 ps, for the $\alpha$, $\beta$, and $\gamma$ sheets, respectively. 

%
\subsection{Method 2: Direct Fitting}
Next we extract \tauq by directly fiting the oscillatory signal to with \autoref{eq:oscfulls}. This method ignores beats in the data but provides a good estimate of \tauq if the field range is large enough.

With this method the quantum lifetimes we obtained for $\alpha$, $\beta$, and $\gamma$ sheets are 1.11 ps, 0.65 ps, and 0.64 ps, respectively. 


\section{Calculation of the Residual Resistivity}

\subsection{The Chambers Formula}

The Boltzmann transport equation can be solved in the semiclassical limit, and within the relaxation-time approximation, for any component of the conductivity tensor. The approach most suitable for calculating the conductivity for arbitrary band structures was developed by Chambers \cite{chambers1952kinetic}, with the conductivity tensor given as 
\begin{equation}
\sigma_{ij}  = \frac{e^2}{4 \pi^3}\!\int \!d^3\bm{k}\left(-\frac{df_0}{d\epsilon}\right)v_i\!\left(\bm{k}\right)v_j\!\left(\bm{k}\right)\tau_t,
\label{eq:chambS}
\end{equation}
where $\int \!d^3\bm{k}$ is an integral over the entire Brillouin zone, $\left(-\frac{df_0}{d\epsilon}\right)$ is the derivative with respect to energy of the equilibrium Fermi distribution function, $v_i\!\left(\bm{k}\right)$ is the $i^{\mathrm{th}}$ component of the quasiparticle velocity at momentum point $\bm{k}$, and $\tau_t$ is the quasiparticle transport lifetime. The velocity is calculated from a tight binding model of the bandstructure using $\bm{v} = \frac{1}{\hbar}\vec{\nabla}_{\bm{k}}\epsilon(\bm{k})$.

In the limit that the temperature is much smaller than any of the hopping parameters in $\epsilon(\bm{k})$, the factor $\left(-\frac{df_0}{d\epsilon}\right)$ can be accurately approximated as a delta function at the Fermi energy. This delta function transforms the integral over the Brillouin zone into an integral over the Fermi surface and introduces a factor of $1/|\vec{\nabla}_{\bm{k}}\epsilon(\bm{k})|$, which is the density of states. \autoref{eq:chambS} is solved numerically by discretizing the Fermi surface and summing $v_i\!\left(\bm{k}\right)v_j\!\left(\bm{k}\right)$ for each point, weighted by the local density of states. We use it to calculate $\sigma_{xx}$ and $\sigma_{xy}$, and use these quantities to calculate $\rho_{xx}=\frac{\sigma_{xx}}{\sigma_{xx}\sigma_{yy}-\sigma_{xy}\sigma_{yx}}$.

\subsection{\sro Band structure}
 We use a standard three-orbital tight-binding model to describe the band structure of \sro \cite{burganov2016strain}. The energy bands are the eigenvalues of the Hamiltonian matrix
 \begin{equation}
\bm{H} = 
\begin{pmatrix}
\epsilon_{xy} & 0 & 0 \\
0 & \epsilon_{xz} & V \\
0 & V & \epsilon_{yz}
\end{pmatrix},
\label{eq:band}
\end{equation}
where $\epsilon_{xy}=-\mu_1-2t_1(\cos(k_xa)+\cos(k_ya))-4t_4\cos(k_xa)\cos(k_ya)$ is the tight binding model of the ruthenium $d_{xy}$ orbitals, $\epsilon_{\left\{x,y\right\}z}=-\mu_2-2t_2cos(k_{\left\{x,y\right\}}a)-2t_3\cos(k_{\left\{y,x\right\}}a)$ are the tight binding models of the ruthenium $d_{xz}$ and $d_{yz}$ orbitals, and $V = 4t_5\sin(k_xa)\sin(ky_a)$ is the hybridization matrix element between the $d_{xz}$ and $d_{yz}$ orbitals. 

We start with tight binding parameters for bulk \sro, given in \citet{burganov2016strain}, and adjust $\mu_1/t_1$ and $\mu_2/t_2$ so that the Fermi surface areas produced by the tight-binding model agree with the areas we measure from the quantum oscillations. We further adjust the bandwidth of each band that results from diagonalizing \autoref{eq:band} so that the effective masses given by the tight binding model also agree with the experimental results. The effective masses for each of the three Fermi surfaces given by \autoref{eq:band} are calculated numerically via
\begin{equation}
m_i = \frac{\hbar^2}{2\pi}\left(\frac{\partial A_k}{\partial \epsilon_i}\right),
\label{eq:}
\end{equation} where $A_k$ is the Fermi surface area in momentum space \cite{shoenberg1984magnetic}. The renormalizations of the bandwidths required to obtain agreement between the tight binding and experimental masses are 0.60, 0.78, and 0.65 for the $\alpha$, $\beta$ and $\gamma$ sheets, respectively. While these scaling factors change the bandwidths from those experimentally determined by ARPES, only the slopes of the bands immediately at the Fermi energy are relevant for quantum oscillations and the electrical resistivity. These renormalizations may account for interaction effects that are not resolvable by ARPES, for example. 

\autoref{tab:compTB} compares the quantum oscillation frequencies and effective masses as given by the tight binding fit of the ARPES spectra on bulk \sro from \citet{burganov2016strain}, the values obtained by scaling that band structure, and the experimental values we obtain from the quantum oscillations.  

\begin{table}[h!]
  \begin{center}
    \begin{tabular}{|l| l| l| l| l| l| l |l |} 
    \hline
         & $F_\alpha$ (T) & $F_\beta$ (T)  & $F_\gamma$ (T) & $m_\alpha$ $(m_e)$ &  $m_\beta$ $(m_e)$ &  $m_\gamma$ $(m_e)$ \\ \hline
       ARPES fit  & 2959  & 12498  & 17886 & 2.0 & 5.7& 11.6\\ \hline
       Scaled & 2988 & 12357 & 18268&3.6&6.4&15.1\\ \hline
      This experiment & 3079 & 12510 & 18259 &3.5&6.3&15.2\\ \hline
    \end{tabular}
  \end{center}
 \caption{Comparison of Fermi surface parameters as given by the tight-binding fit in \citet{burganov2016strain}, the values obtained from scaling that tight binding model, and the experimental values from this work.}
    \label{tab:compTB} 
\end{table}

\subsection{Finite element simulation of the contact geometry}
We analyze the contact geometry of our device, shown in Fig. 1a, in order to determine the absolute resistivity (and thus transport mean free path) with high precision. Typically, when the width of the voltage contacts is comparable to the distance between the contacts, there is large uncertainty in the conversion from resistance to resistivity. With the help of a finite element simulation, we can determine the correct geometric factor for our sample and thus reduce the uncertainty substantially. 

\begin{figure}[h!]
\includegraphics[width=1\columnwidth]{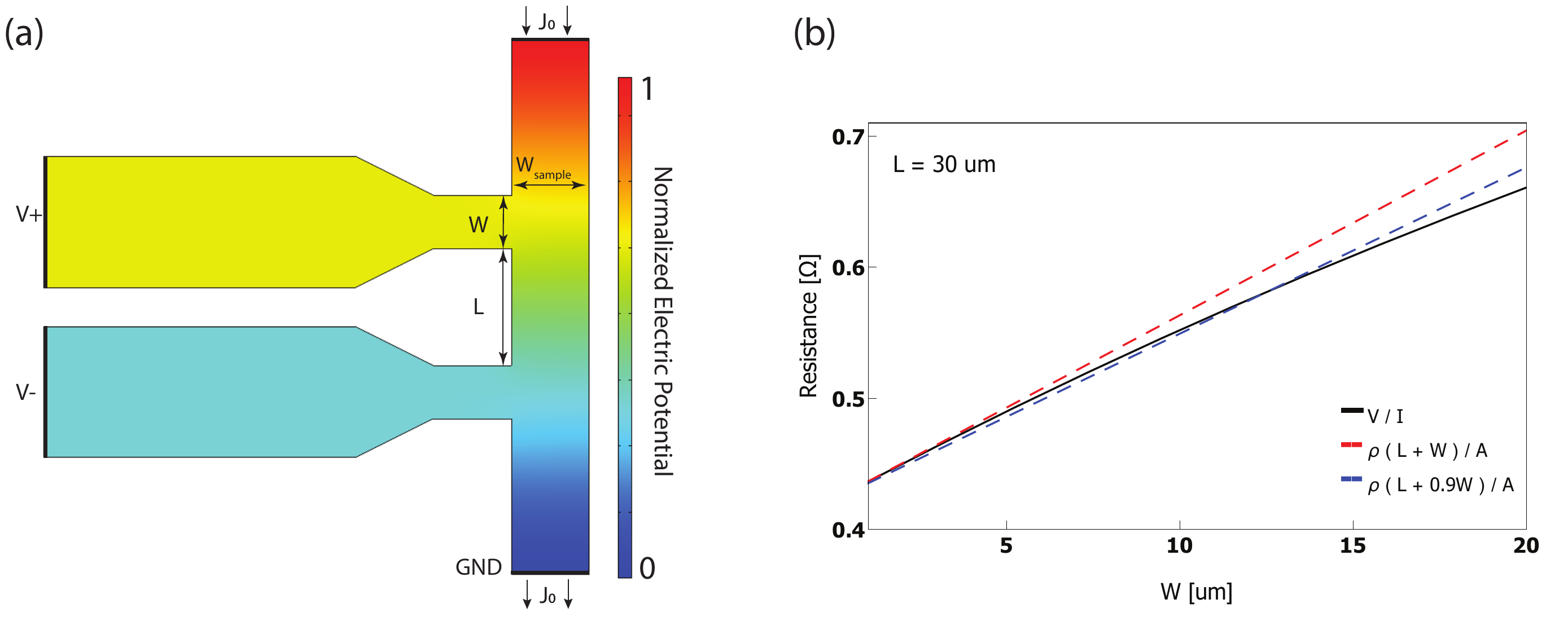}%
\caption{\textbf{Finite element simulation on the effect of the width of voltage contact.}
(a) The model used in the finite element simulation. $\rm{W_{\rm sample}}$ is set to 20 $\mu$m and L is set to 30 $\mu$m, which are the dimensions of the device. The color scale represents the normalized electric potential. (b) Resistance versus voltage pad width W, with all other dimensions and the resistivity of the material fixed. The black line represents the measured resistance and the red and blue dashed line represents the calculated resistance with $l$ set to L + W and L + 0.9W, respectively.}
\label{fig:comsol}
\end{figure}

The model we use in the simulation is shown in \autoref{fig:comsol}. We fix the current density flowing into and out of the sample, normal to the top edge and the bottom edge, and solve for the stationary solution. We obtain the measured resistance using $R = (V_{+}-V_{-})/(J_0*\rm{W_{sample}*t})$, where $V_{+}$ and $V_{-}$ are the average electric potential along the edges labeled $V_{+}$ and $V_{-}$ in \autoref{fig:comsol}a, $J_0$ is the current density, $\rm{W_{sample}}$ is the sample width, and $t$ is the film thickness. As the resistivity of the material, $\rho$, is fixed in the model, we can also calculate a value for the resistance using $R = \rho * l / \rm{A}$, where $l$ is some effective sample length, and $\rm{A} = \rm{t * W_{sample}}$ is the cross section area. By varying the effective sample length and comparing the calculated resistance with the resistance measured via $V_{+}$ and $V_{-}$, we can determine the correct effective sample length. 

We set $L$ to 30 $\mu$m, $\rm{W_{\rm sample}}$ to 20 $\mu$m, and T to be 100 nm, which are the dimensions of our device. \autoref{fig:comsol}b shows how the measured (V/I) and calculated resistances change with contact width W. At our experimental contact width of W $=10~\mu$m, we find that an effective sample length $l$ of $\rm{L} + 0.9*\rm{W}$ agrees with the measured resistance to better than 1\%. 

\section{Method for obtaining the value and estimation of the uncertainties in Table I }

The frequencies and their uncertainties for the three pockets in thin film \sro are obtained from the Fourier transform shown in Fig. 2b. The frequency for $\alpha$ is taken to be the average of the two peaks at $3\alpha$, divided by three, and the uncertainty is taken to be half of their difference, divided by three. This avoids interference effects due to multiple closely-spaced frequencies that arise from inter-layer coupling (such interference is likely shifting the first harmonic peak down in frequency). The frequency and uncertainty in frequency of $\beta$ and $\gamma$ pockets are obtained by fitting a lorentzian function to the data near the corresponding FFT peaks. 

The frequencies and their uncertainties for the three pockets in single crystal \sro are taken to be the average and half of their difference between the values reported in two de Haas-van Alphen studies \cite{bergemann1999quantum,mackenzie1996quantum}. The same procedure is used to obtain the effective masses and their uncertainties.

\section{Method for obtaining the value of $\tau_t/\tau_q$ and its uncertainty}

We start by evaluating the uncertainty in $\tau_t / \tau_q$ that comes from the uncertainties in $\tau_q.$ We use the tight-binding parameters that best match the Fermi surface areas and cyclotron masses we measure (see the previous section). With the adjusted tight-binding parameters and with the measured quantum lifetimes for the three pockets, we solve the Boltzmann equation numerically and obtain the resistivity at zero magnetic field. We then have only one parameter, $\tau_t / \tau_q$, to determine. For fixed values of the three $\tau_q$’s, we vary the ratio of $\tau_t / \tau_q$ to minimize the difference between the calculated resistivity and the measured resistivity. We repeat this procedure at the upper and lower bounds of $\tau_q$ for each of the three pockets. This determines the uncertainty in $\tau_t / \tau_q$ from the uncertainties in the quantum lifetimes of $\alpha$, $\beta$, and $\gamma$ to be $\pm0.02$, $\pm0.01$, and $\pm0.04$, respectively. Added in quadrature, this gives a total uncertainty in $\tau_t / \tau_q$ of $\pm0.046$. 

\begin{figure}[h!]
\includegraphics[width=.6\columnwidth]{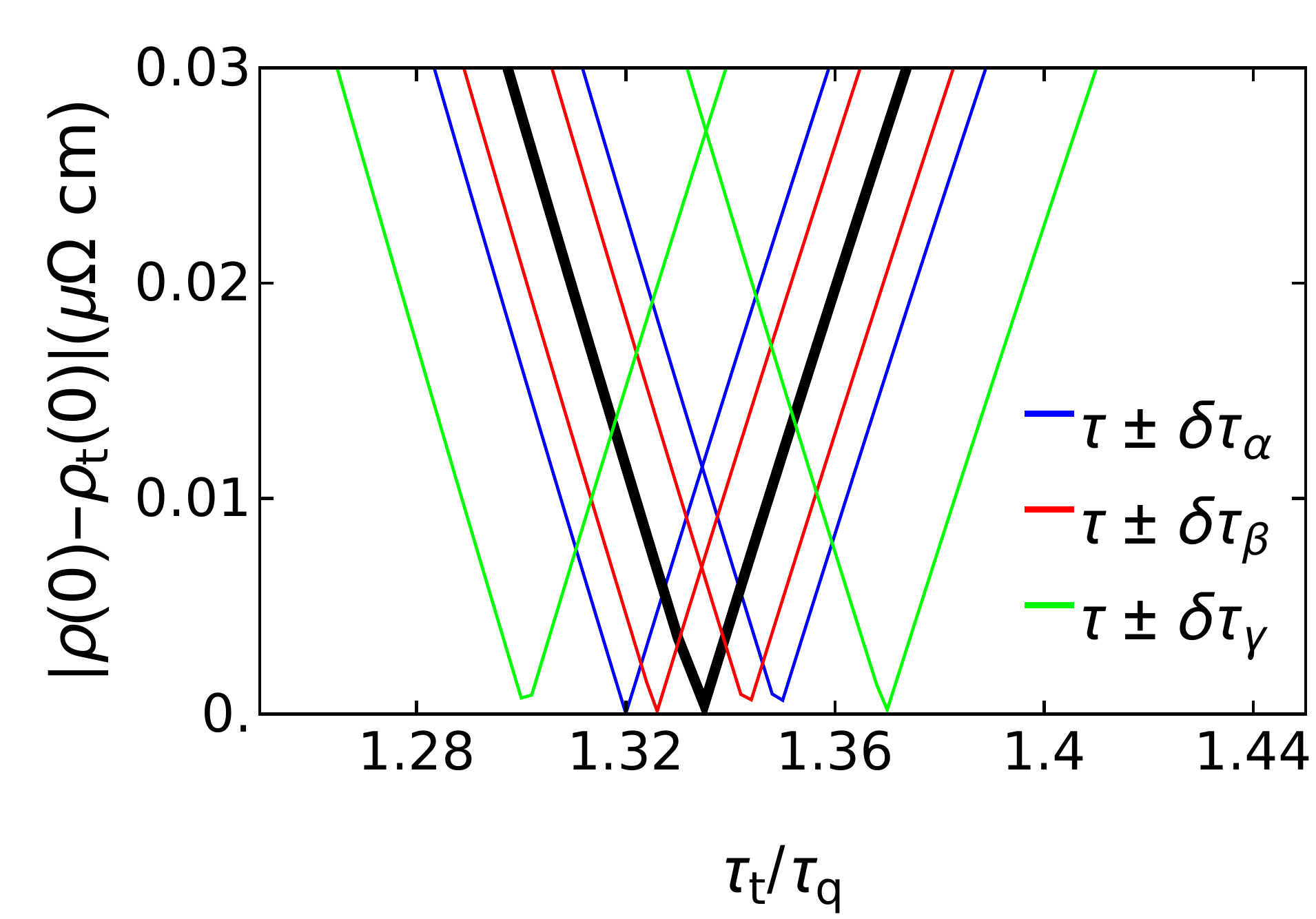}%
\caption{\textbf{Uncertainty estimation for $\protect{\tau_t/\tau_q}$} The absolute value of the difference between the calculated resistivity and the measured residual resistivity as a function of $\tau_t/\tau_q$. The blue, red, and green lines are used to determine the uncertainty from the quantum lifetime of the $\alpha$, $\beta$, and $\gamma$ pockets, respectively. The black line uses the measured quantum lifetimes shown in Table I in the main text.  }
\label{fig:uncertainty}
\end{figure}

Next there is uncertainty in the experimental value of the residual resistivity. This is due to uncertainty in the sample dimensions, as well as uncertainties in the measurement procedure. The uncertainties in the lateral dimensions of the sample come from the resolution of the photolithography process and are $\pm0.5$ um. The thickness of the film is known precisely from the STEM images; the presence of step edges introduces an uncertainty of $\pm1$ nm in the thickness. The total uncertainty in the residual resistivity from sample dimensions is determined to be $\pm 3.7$ \% ($\pm$ 0.037 $\mu\Omega$.cm). The uncertainty in the output of the current source is $\pm 0.1$\% and the uncertainty in the lock-in measurement is $\pm 1$\%. Assuming that all the measurement uncertainties are uncorrelated, we find the total uncertainty in the residual resistivity to be $\pm $3.8 \%, or $\pm 0.038 \mu\Omega$.cm.  Using the slope of $\tau_t / \tau_q$ as a function of the residual resistivity (\autoref{fig:uncertainty}), this produces an uncertainty in $\tau_t / \tau_q$ of $\pm 0.04$.

Finally, combining the uncertainties from $\tau_q$ with the uncertainty from the residual resistivity, the final value of $\tau_t / \tau_q$ is 1.34 ± 0.06. We round this value to 1.3 ± 0.1.

\section{Cross-sectional STEM}

\begin{figure}[h!]
\includegraphics[width=.6\columnwidth]{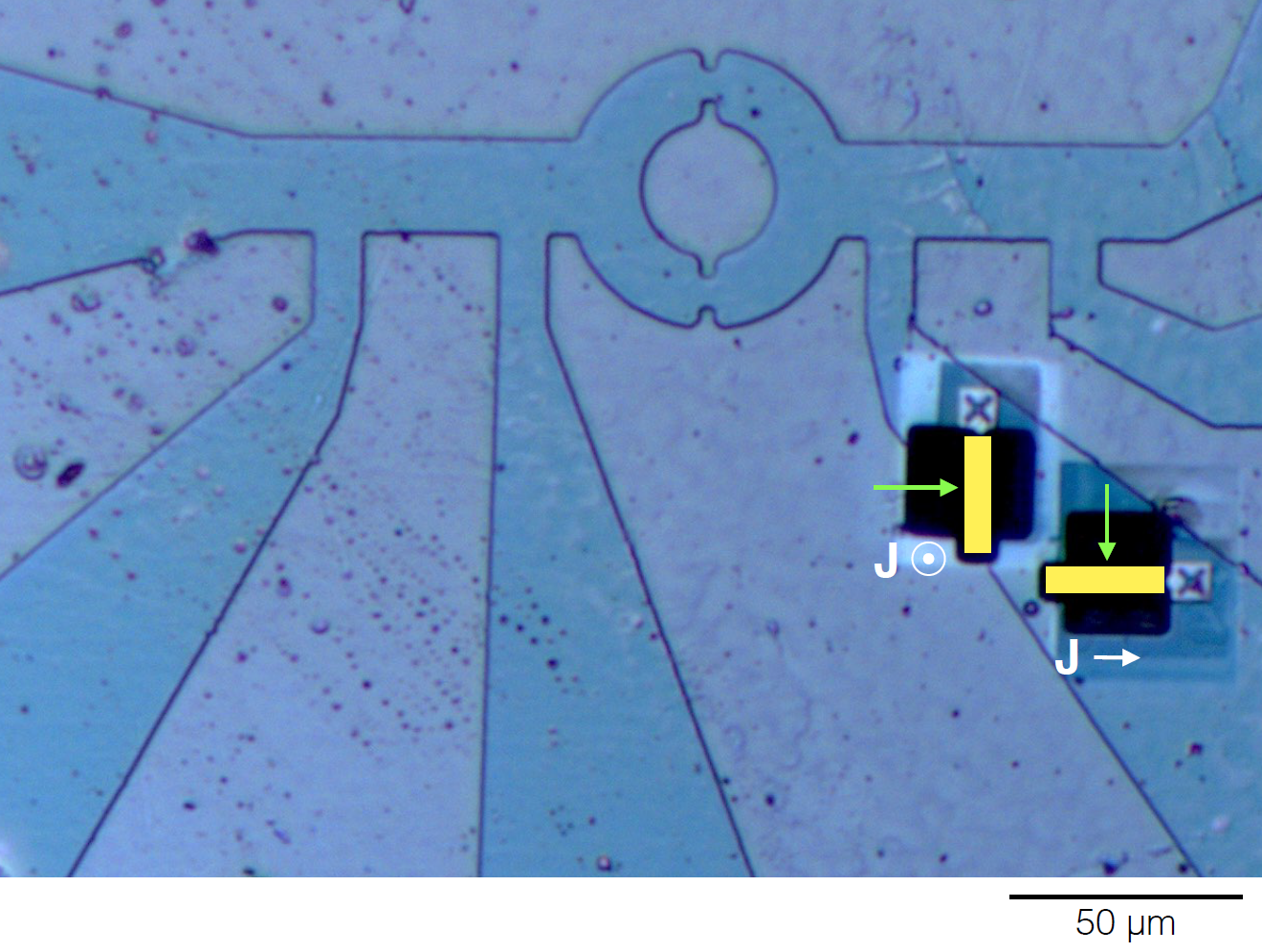}%
\caption{\textbf{Device and STEM specimen locations.} Optical micrograph showing the location and orientation of the extracted cross sections for STEM experiments. The top-view orientation of the lamella cut from each location is indicated by the thick yellow line. The green arrows indicate the projection axes of STEM imaging for each case.}
\label{fig:lamella}
\end{figure}

\begin{figure}[h!]
\includegraphics[width=1\columnwidth]{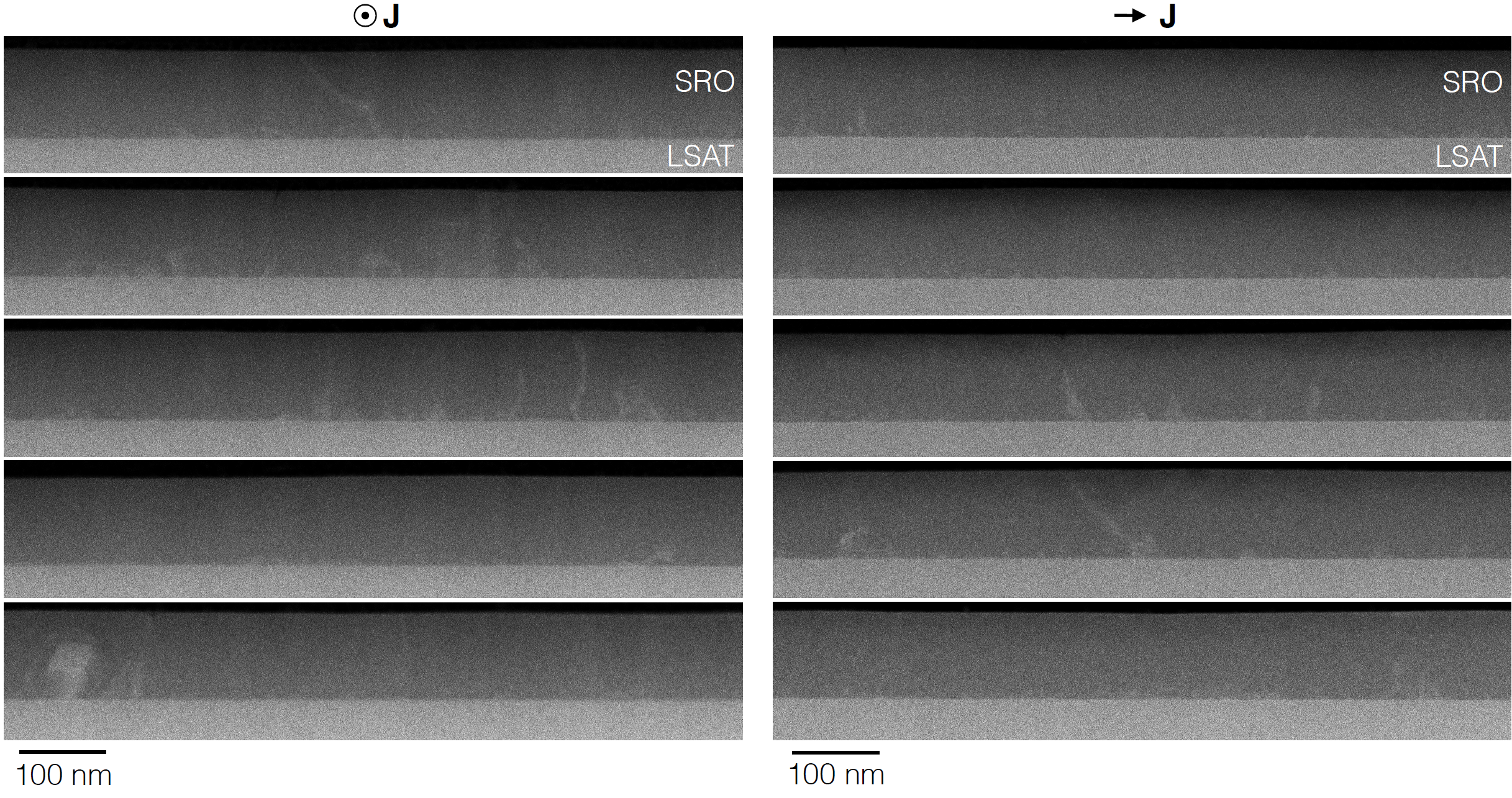}%
\caption{\textbf{Cross-sectional STEM.}
The contribution of small scattering angles in large field-of-view MAADF STEM images of the \sro film in each pseudocubic projection
highlight the extended lattice defects with brighter contrast. Nearly all of the extended defects appear to nucleate at the interface with the
substrate, suggesting that step edges or other aspects of the LSAT surface may play a limiting role in the growth of clean \sro films for
these devices. Some defect structures extending throughout the entire film all the way to the surface are observed in each projection, with a
distribution density on the order one per $\mu$m or fewer. No significant difference in defect density is observed between the two projections.}
\label{fig:stemS1}
\end{figure}

\autoref{fig:lamella} shows where the lamellas were cut for the STEM imaging. \autoref{fig:stemS1} shows several large field-of-view MAADF STEM images for the two lamellas, including the images shown in Fig. 3a and b.

\begin{thebibliography}{25}
\providecommand{\natexlab}[1]{#1}
\providecommand{\url}[1]{\texttt{#1}}
\expandafter\ifx\csname urlstyle\endcsname\relax
  \providecommand{\doi}[1]{doi: #1}\else
  \providecommand{\doi}{doi: \begingroup \urlstyle{rm}\Url}\fi

\bibitem[Pustogow et~al.(2019)Pustogow, Luo, Chronister, Su, Sokolov,
  Jerzembeck, Mackenzie, Hicks, Kikugawa, Raghu, Bauer, and
  Brown]{PustogowNat2019}
 A. Pustogow, Y. Luo, A. Chronister, Y.-S. Su, D. A. Sokolov, F. Jerzembeck, A. P. Mackenzie, C. W. Hicks, N. Kikugawa, S. Raghu, E. D. Bauer, and S. E. Brown, \newblock Constraints on the Superconducting Order Parameter in {Sr\ensuremath{_2}RuO\ensuremath{_4}\xspace} from Oxygen-17 Nuclear Magnetic Resonance, \newblock Nature \textbf{574}, 72 (2019).

\bibitem[Kivelson et~al.(2020)Kivelson, Yuan, Ramshaw, and
  Thomale]{kivelson2020proposal}
 S. A. Kivelson, A. C. Yuan, B. Ramshaw, and R. Thomale, \newblock A Proposal for Reconciling Diverse Experiments on the Superconducting State in {Sr\ensuremath{_2}RuO\ensuremath{_4}\xspace}, \newblock Npj Quantum Mater. \textbf{5}, 43 (2020).


\bibitem[Ghosh et~al.(2021)Ghosh, Shekhter, Jerzembeck, Kikugawa, Sokolov,
  Brando, Mackenzie, Hicks, and Ramshaw]{ghosh2020thermodynamic}
  S. Ghosh, A. Shekhter, F. Jerzembeck, N. Kikugawa, D. A. Sokolov, M. Brando, A. P. Mackenzie, C. W. Hicks, and B. J. Ramshaw, Thermodynamic Evidence for a Two-Component Superconducting Order Parameter in {Sr\ensuremath{_2}RuO\ensuremath{_4}\xspace}, Nat. Phys. \textbf{17}, 199 (2021).

\bibitem[Raghu et~al.(2010)Raghu, Kapitulnik, and Kivelson]{raghu2010hidden}
S. Raghu, A. Kapitulnik, and S. A. Kivelson, Hidden Quasi-One-Dimensional Superconductivity in {Sr\ensuremath{_2}RuO\ensuremath{_4}\xspace}, Phys. Rev. Lett. \textbf{105}, 136401 (2010).

\bibitem[Zhang et~al.(2018)Zhang, Huang, Yang, and Yao]{zhang2018superconducting}
L.-D. Zhang, W. Huang, F. Yang, and H. Yao, Superconducting Pairing in {Sr\ensuremath{_2}RuO\ensuremath{_4}\xspace} from Weak to Intermediate Coupling, Phys. Rev. B \textbf{97}, 060510 (2018).

\bibitem[Hicks et~al.(2014)Hicks, Brodsky, Yelland, Gibbs, Bruin, Barber,
  Edkins, Nishimura, Yonezawa, Maeno, et~al.]{hicks2014strong}
C. W. Hicks, D. O. Brodsky, E. A. Yelland, A. S. Gibbs, J. A. N. Bruin, M. E. Barber, S. D. Edkins, K. Nishimura, S. Yonezawa, Y. Maeno, and A. P. Mackenzie, Strong Increase of Tc of {Sr\ensuremath{_2}RuO\ensuremath{_4}\xspace} Under Both Tensile and Compressive Strain, Science \textbf{344}, 283 (2014).

\bibitem[Burganov et~al.(2016)Burganov, Adamo, Mulder, Uchida, King, Harter,
  Shai, Gibbs, Mackenzie, Uecker, et~al.]{burganov2016strain}
 B. Burganov, C. Adamo, A. Mulder, M. Uchida, P. D. C. King, J. W. Harter, D. E. Shai, A. S. Gibbs, A. P. Mackenzie, R. Uecker, M. Bruetzam, M. R. Beasley, C. J. Fennie, D. G. Schlom, and K. M. Shen, Strain Control of Fermiology and Many-Body Interactions in Two-Dimensional Ruthenates, Phys. Rev. Lett. \textbf{116}, 197003 (2016).


\bibitem[Madhavan et~al.({1996})Madhavan, Schlom, Dabkowski, Dabkowska, and
  Liu]{madhavan1996a}
 S. Madhavan, D. G. Schlom, A. Dabkowski, H. A. Dabkowska, and Y. Liu, Growth of Epitaxial A‐axis and C‐axis Oriented {Sr\ensuremath{_2}RuO\ensuremath{_4}\xspace} Films, Appl. Phys. Lett. \textbf{68}, 559 (1996).

\bibitem[Mackenzie et~al.(1998)Mackenzie, Haselwimmer, Tyler, Lonzarich, Mori,
  Nishizaki, and Maeno]{MackenziePRL1998}
  A. P. Mackenzie, R. K. W. Haselwimmer, A. W. Tyler, G. G. Lonzarich, Y. Mori, S. Nishizaki, and Y. Maeno, Extremely Strong Dependence of Superconductivity on Disorder in {Sr\ensuremath{_2}RuO\ensuremath{_4}\xspace}, Phys. Rev. Lett. \textbf{80}, 161 (1998).

\bibitem[Nair et~al.(2018)Nair, Ruf, Schreiber, Miao, Grandon, Baek, Goodge,
  Ruff, Kourkoutis, Shen, et~al.]{nair2018demystifying}
H. P. Nair, J. P. Ruf, N. J. Schreiber, L. Miao, M. L. Grandon, D. J. Baek, B. H. Goodge, J. P. C. Ruff, L. F. Kourkoutis, K. M. Shen, and D. G. Schlom, Demystifying the Growth of Superconducting {Sr\ensuremath{_2}RuO\ensuremath{_4}\xspace} Thin Films, APL Materials \textbf{6}, 101108 (2018).
\bibitem[Mark et~at.(2001)]{mark2001suppression}
M. A. Zurbuchen, Y. Jia, S. Knapp, A. H. Carim, D. G. Schlom, L.-N. Zou, and Y. Liu, Suppression of Superconductivity by Crystallographic Defects in Epitaxial S${\mathrm{r}}_{2}$Ru${\mathrm{O}}_{4}$ Films, Appl. Phys. Lett. \textbf{78}, 2351 (2001).


\bibitem[Mark et~al.(2003)]{mark2003defect}
M. A. Zurbuchen, Y. Jia, S. Knapp, A. H. Carim, D. G. Schlom, and X. Q. Pan, Defect Generation by Preferred Nucleation in Epitaxial S${\mathrm{r}}_{2}$Ru${\mathrm{O}}_{4}$/LaAl${\mathrm{O}}_{4}$, Appl. Phys. Lett. \textbf{83}, 3891 (2003).

\bibitem[Uchida et~al.(2017)Uchida, Ide, Watanabe, Takahashi, Tokura, and
  Kawasaki]{uchida2017molecular}
M. Uchida, M. Ide, H. Watanabe, K. S. Takahashi, Y. Tokura, and M. Kawasaki, Molecular Beam Epitaxy Growth of Superconducting {Sr\ensuremath{_2}RuO\ensuremath{_4}\xspace} Films, APL Materials \textbf{5}, 106108 (2017).

\bibitem[Garcia et~al.(2020)Garcia, Di~Bernardo, Kimbell, Vickers, Massabuau,
  Komori, Divitini, Yasui, Lee, Kim, et~al.]{garcia2020pair}
C. M. P. Garcia, A. Di Bernardo, G. Kimbell, M. E. Vickers, F. C.-P. Massabuau, S. Komori, G. Divitini, Y. Yasui, H. G. Lee, J. Kim, B. Kim, M. G. Blamire, A. Vecchione, R. Fittipaldi, Y. Maeno, T. W. Noh, and J. W. A. Robinson, Pair Suppression Caused by Mosaic-Twist Defects in Superconducting {Sr\ensuremath{_2}RuO\ensuremath{_4}\xspace} Thin-Films Prepared Using Pulsed Laser Deposition, Commun Mater \textbf{1}, 1 (2020).


\bibitem[Shoenberg(1984)]{shoenberg1984magnetic}
D. Shoenberg, Magnetic Oscillations in Metals (Cambridge University Press, Cambridge Cambridgeshire; New York, 1984). 

\bibitem[Bergemann et~al.(2003)Bergemann, Mackenzie, Julian, Forsythe, and
  Ohmichi]{bergemann2003quasi}
C. Bergemann, A. P. Mackenzie, S. R. Julian, D. Forsythe, and E. Ohmichi, Quasi-Two-Dimensional Fermi Liquid Properties of the Unconventional Superconductor {Sr\ensuremath{_2}RuO\ensuremath{_4}\xspace}, Advances in Physics \textbf{52}, 639 (2003).

\bibitem[Bergemann et~al.(1999)Bergemann, Julian, Mackenzie, Tyler, Farrell,
  Maeno, and Nishizaki]{bergemann1999quantum}
C. Bergemann, S. R. Julian, A. P. Mackenzie, A. W. Tyler, D. E. Farrell, Y. Maeno, and S. Nishizaki, Quantum Oscillations and Overcritical Torque Interaction in {Sr\ensuremath{_2}RuO\ensuremath{_4}\xspace}, Physica C: Superconductivity \textbf{317–318}, 444 (1999).

\bibitem[Bergemann et~al.(2001)Bergemann, Brooks, Balicas, Mackenzie, Julian,
  Mao, and Maeno]{bergemann2001normal}
  C. Bergemann, J. S. Brooks, L. Balicas, A. P. Mackenzie, S. R. Julian, Z. Q. Mao, and Y. Maeno, Normal State of the Unconventional Superconductor {Sr\ensuremath{_2}RuO\ensuremath{_4}\xspace} in High Magnetic Fields, Physica B: Condensed Matter \textbf{294–295}, 371 (2001).

\bibitem[Mackenzie et~al.(1996{\natexlab{a}})Mackenzie, Julian, Diver,
  McMullan, Ray, Lonzarich, Maeno, Nishizaki, and Fujita]{mackenzie1996quantum}
A. P. Mackenzie, S. R. Julian, A. J. Diver, G. J. McMullan, M. P. Ray, G. G. Lonzarich, Y. Maeno, S. Nishizaki, and T. Fujita, Quantum Oscillations in the Layered Perovskite Superconductor S${\mathrm{r}}_{2}$Ru${\mathrm{O}}_{4}$, Phys. Rev. Lett. \textbf{76}, 3786 (1996).


\bibitem[Chambers(1952)]{chambers1952kinetic}
R. G. Chambers, The Kinetic Formulation of Conduction Problems, Proc. Phys. Soc. A \textbf{65}, 458 (1952).

\bibitem[Mackenzie et~al.(1996{\natexlab{b}})Mackenzie, Julian, Diver,
  Lonzarich, Hussey, Maeno, Nishizaki, and Fujita]{mackenzie_calculation_1996}
A. P. Mackenzie, S. R. Julian, A. J. Diver, G. G. Lonzarich, N. E. Hussey, Y. Maeno, S. Nishizaki, and T. Fujita, Calculation of Thermodynamic and Transport Properties of Sr2RuO4 at Low Temperatures Using Known Fermi Surface Parameters, Physica C: Superconductivity \textbf{263}, 510 (1996).

\bibitem[Breznay et~al.(2016)Breznay, Hayes, Ramshaw, McDonald, Krockenberger,
  Ikeda, Irie, Yamamoto, and Analytis]{breznay2016shubnikov}
N. P. Breznay, I. M. Hayes, B. J. Ramshaw, R. D. McDonald, Y. Krockenberger, A. Ikeda, H. Irie, H. Yamamoto, and J. G. Analytis, Shubnikov-de Haas Quantum Oscillations Reveal a Reconstructed Fermi Surface near Optimal Doping in a Thin Film of the Cuprate Superconductor ${\mathrm{Pr}}_{1.86}{\mathrm{Ce}}_{0.14}{\mathrm{CuO}}_{4\ifmmode\pm\else\textpm\fi{}\ensuremath{\delta}}$, Phys. Rev. B \textbf{94}, 104514 (2016).


\bibitem[Sharma et~al.(2020)Sharma, Edkins, Wang, Kostin, Sow, Maeno,
  Mackenzie, Davis, and Madhavan]{sharma2019momentum}
R. Sharma, S. D. Edkins, Z. Wang, A. Kostin, C. Sow, Y. Maeno, A. P. Mackenzie, J. C. S. Davis, and V. Madhavan, Momentum-Resolved Superconducting Energy Gaps of S${\mathrm{r}}_{2}$Ru${\mathrm{O}}_{4}$ from Quasiparticle Interference Imaging, PNAS \textbf{117}, 5222 (2020).

\bibitem[Manfra et~al.(2004)Manfra, Simon, Baldwin, Sergent, West, Molnar, and
  Caissie]{manfra_quantum_2004}
M. J. Manfra, S. H. Simon, K. W. Baldwin, A. M. Sergent, K. W. West, R. J. Molnar, and J. Caissie, Quantum and Transport Lifetimes in a Tunable Low-Density AlGaN/GaN Two-Dimensional Electron Gas, Appl. Phys. Lett. \textbf{85}, 5278 (2004).

\bibitem[Schlom et~al.(2014)Schlom, Chen, Fennie, Gopalan, Muller, Pan, Ramesh,
  and Uecker]{schlom2014elastic}
D. G. Schlom, L.-Q. Chen, C. J. Fennie, V. Gopalan, D. A. Muller, X. Pan, R. Ramesh, and R. Uecker, Elastic Strain Engineering of Ferroic Oxides, MRS Bulletin \textbf{39}, 118 (2014).

\end{thebibliography}
\end{document}